# Parameterized Complexity on a New Sorting Algorithm: A Study in Simulation

[1]Prashant Kumar, [2]Anchala Kumari and [3]Soubhik Chakraborty*

[1,3] Department of Applied Mathematics  
Birla Institute of Technology, Mesra  
Ranchi – 835215, India

[2]Department of Statistics  
Patna University  
Patna – 800005, India

prashant.mtech@gmail.com, anchalak@yahoo.com, soubhikc@yahoo.co.in

**ABSTRACT.** Sundararajan and Chakraborty (2007) introduced a new sorting algorithm by modifying the fast and popular Quick sort and removing the interchanges. In a subsequent empirical study, Sourabh, Sundararajan and Chakraborty (2007) demonstrated that this algorithm sorts inputs from certain probability distributions faster than others and the authors made a list of some standard probability distributions in decreasing order of speed, namely, Continuous uniform < Discrete uniform < Binomial < Negative Binomial < Poisson < Geometric < Exponential < Standard Normal. It is clear from this interesting second study that the algorithm is sensitive to input probability distribution. Based on these pervious findings, in the present paper we are motivated to do some further study on this sorting algorithm through simulation and determine the appropriate empirical model which explains its average sorting time with special emphasis on parameterized complexity.

**KEYWORDS.** New Sorting Algorithm; average complexity; comparisons; interchanges; simulation; parameterized complexity; pivot; temporary array.

## 1. Introduction

In many if not all spheres of human activities we need to keep information as well as retrieve the same, so we must keep it in some sensible order. Computers spend a considerable amount of their time keeping data in order.

The objective of the sorting method is to rearrange the data so that their keys are ordered according to some well-defined ordering rules. The essence of sorting is a mapping relationship between data values and their corresponding ordered positions. A perfect sorting algorithm will make us accomplish our goal via just one calculation, substituting the value of elements into the function and returning us their location. Among the algorithms that are frequently used are the sorting algorithms which find a number of applications in both commercial and scientific disciplines. For a sound literature on sorting, see Knuth (2000). For sorting algorithms with special reference to input probability distributions, see Mahmoud (2000).
This paper describes a new sorting algorithm and its dependence to parameter(s) of certain probability distributions characterizing the sorting elements through simulation.

Sundararajan and Chakraborty (2007) introduced a new sorting algorithm which can accomplish sorting without interchanges. In order to make this paper self contained, we are providing it again:

Step 1: Initialize the first element of the array as a pivot element .

Step 2: Starting from the second element , compare it to the pivot element .

    Step 2.1: If pivot < element then place the element in the last unfilled position of the temporary array ( of same size as the original one).

    Step 2.2: If pivot >=element then place the element in the first unfilled position of the temporary array.

Step 3: Repeat step 2 till the last element of the array.

Step 4: Finally place the pivot element in the blank position of the temporary array.
(Remark: the blank position is created because one element of the original array was taken out as pivot)

Step 5: Split the array into two, based on the pivot element's position.

Step 6: Repeat steps 1-5 till the array is sorted completely.

It turns out that the basic mechanism of this algorithm is similar to that of Quick sort except for removal of interchanges and therefore both algorithms have O(nlogn) complexity theoretically in the average situation as it depends on the number of comparisons and not interchanges. In a subsequent empirical study, Sourabh, Sundararajan and Chakraborty (2007) demonstrated that this algorithm sorts inputs from certain probability distributions faster than those from others and the authors made a list of some standard probability distributions in decreasing order of speed, namely, Continuous uniform < Discrete Uniform < Binomial < Negative Binomial < Poisson < Geometric < Exponential < Standard Normal. It is clear from this interesting second study that the algorithm is sensitive to input probability distribution. Based on these pervious findings, in the present paper we are motivated to do some further study on this sorting algorithm through simulation and determine the appropriate empirical model which explains its average sorting time with special emphasis on parameterized complexity.

To make the experiment more accurate and reliable, we went for higher n and used high specific software tool Borland Turbo c++ version 5.02. We observed that if parameters are changed then average time complexity (mean elapsed time) of the new sorting algorithm also changed. This algorithm works on the divide & conquer strategy similar to Quick sort except removal of interchanges. As mentioned earlier, both Quick sort and New Sort have same average time complexity $O(n \log n)$ and same worst case $O(n^2)$ complexity as this is decided by comparisons rather than interchanges.

## 2. Empirical results and discussion

Our interest is to observe how our sorting algorithm depends on the parameter(s) of the input distribution. Because the Quick Sort has been found to be sensitive to parameters of input probability distributions and the New sorting algorithm is a modification of Quick sort we intuitively guessed that the New sorting algorithm might exhibit such sensitivity. For this we studied the behaviour of the new sorting algorithm by fixing the input size and varying parameter of specific input probability distribution. We simulated different probability distributions and tested the above.

Earlier empirical results show sensitiveness of different probability distributions according to size of input (Sourabh, Sundararajan and Chakraborty 2007). According to those empirical results and graph the sorting algorithms sorts inputs from a continuous uniform distribution faster than those from most other standard distributions.

Now we are presenting some empirical results of different probability distributions by conducting computer experiments with time as response. For the link between algorithmic complexity and computer experiments see [Chakraborty, Sourabh, Bose and Sushant (2007)]. [Chakraborty and Sourabh(2007)] and [Chakraborty, Wahi, Sourabh, Saraswati and Mukherjee (2007)] show why it makes sense to work directly on time if average case complexity is in question. The present empirical results of different distribution show how the average complexity of the sorting algorithms depends on parameters of input distributions. For simulating variates from different distributions the respective algorithms were taken from [Ross, (2006)] and varying the parameters of different distribution, for a given n, we sorted the array of size n by new sorting algorithm in **Borland Turbo C ++ version 5.02** software. We would like to mention here that although our intention was not to vary n but parameters of input distributions only, we were forced to take readings at appropriate distinct values of n (although fixed for a particular distribution input but varying for another) because our software did not support the program run and came out with some error message.

**(a) Discrete Uniform Distribution:**

The discrete uniform distribution depends on the parameter K [1….K] . In this distribution we vary the range of K, i.e., we are varying K from 5 to 50 in the interval of 5. And we fixed the array size of n = 20,000. Then we get following empirical result which showed in table 8.1.

**Table 1. Mean sort time in sec versus K for discrete U[1, 2..K] input when n = 20,000**

| K | $Trail_1$ | $Trail_2$ | $Trail_3$ | $Trail_4$ | $Trail_5$ | $Trail_6$ | $Trail_7$ | $Trail_8$ | $Trail_9$ | $Trail_{10}$ | Mean |
|---|---|---|---|---|---|---|---|---|---|---|---|
| 5 | 2.219 | 3.421 | 3.422 | 3.078 | 2.813 | 2.703 | 2.532 | 2.094 | 2.109 | 2.079 | 2.647 |
| 10 | 1.078 | 1.187 | 1.265 | 1.344 | 1.188 | 1.157 | 1.484 | 1.282 | 1.219 | 1.235 | 1.244 |
| 15 | 0.703 | 0.671 | 0.750 | 0.672 | 0.688 | 0.718 | 0.782 | 0.672 | 0.704 | 0.687 | 0.7047 |
| 20 | 0.547 | 0.5 | 0.516 | 0.515 | 0.563 | 0.578 | 0.625 | 0.813 | 0.641 | 0.781 | 0.6079 |
| 25 | 0.406 | 0.422 | 0.407 | 0.485 | 0.437 | 0.469 | 0.391 | 0.453 | 0.39 | 0.438 | 0.4298 |
| 30 | 0.469 | 0.563 | 0.547 | 0.453 | 0.562 | 0.531 | 0.422 | 0.562 | 0.406 | 0.390 | 0.4905 |
| 35 | 0.297 | 0.281 | 0.312 | 0.296 | 0.344 | 0.313 | 0.3590 | 0.282 | 0.343 | 0.297 | 0.3124 |
| 40 | 0.265 | 0.266 | 0.250 | 0.329 | 0.281 | 0.344 | 0.313 | 0.297 | 0.328 | 0.359 | 0.3032 |
| 45 | 0.235 | 0.281 | 0.234 | 0.297 | 0.219 | 0.297 | 0.218 | 0.375 | 0.390 | 0.360 | 0.239 |
| 50 | 0.281 | 0.344 | 0.250 | 0.203 | 0.265 | 0.343 | 0.266 | 0.297 | 0.329 | 0.312 | 0.289 |

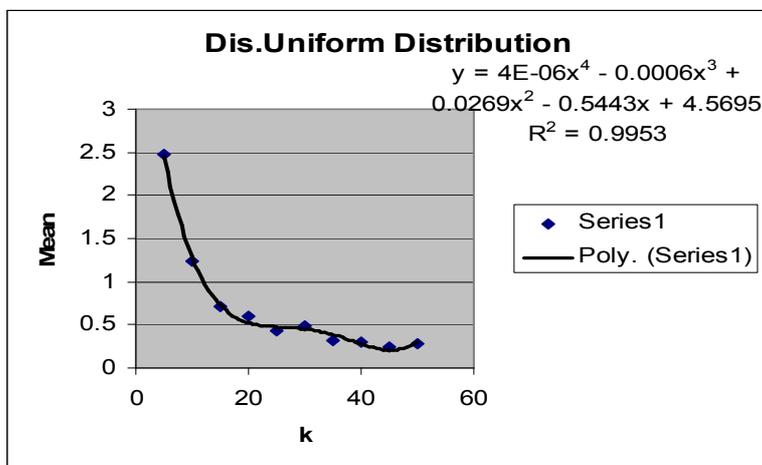

**Figure 1. Mean time in sec versus K corresponding to table 1**

**Result**: Above data is best explained by a polynomial of order four and $R^2=0.9953$. Average sorting time clearly depends on the parameter K of discrete uniform[1,2…..K] input.

**(b) Poisson Distribution:**

The Poisson distribution depends on the parameter $\lambda$. Lambda (which is both the mean and the variance) should be not large as this is the distribution of rare events. In this distribution we varied $\lambda$ from 1 to 5.5 in the interval of 0.5. And we fixed the array size of n = 50,000.

**Table 2. Mean sort time in sec versus Lambda ($\lambda$) for Poisson, when n =50,000**

| $\lambda$ | $TRAIL_1$ | $T_2$ | $T_3$ | $T_4$ | $T_5$ | $T_6$ | $T_7$ | $T_8$ | $T_9$ | $T_{10}$ | MEAN |
|---|---|---|---|---|---|---|---|---|---|---|---|
| 1.0 | 3.219 | 3.313 | 3.344 | 3.250 | 3.422 | 3.406 | 3.360 | 3.297 | 3.531 | 3.234 | 3.3346 |
| 1.5 | 3.442 | 3.000 | 2.829 | 4.250 | 2.562 | 2.578 | 2.953 | 2.906 | 2.562 | 2.812 | 2.9894 |
| 2.0 | 2.344 | 2.188 | 2.353 | 2.422 | 2.219 | 2.313 | 2.218 | 2.250 | 2.391 | 2.203 | 2.2907 |
| 2.5 | 2.109 | 1.922 | 2.031 | 1.969 | 2.047 | 2.234 | 2.641 | 2.063 | 2.515 | 3.188 | 2.2719 |
| 3.0 | 1.750 | 2.015 | 1.828 | 2.031 | 1.937 | 1.719 | 1.875 | 2.469 | 2.125 | 2.453 | 2.0202 |
| 3.5 | 2.062 | 1.609 | 1.812 | 1.906 | 1.750 | 1.640 | 1.719 | 1.687 | 1.890 | 1.797 | 1.7872 |
| 4.0 | 1.594 | 1.766 | 1.641 | 1.857 | 1.547 | 1.625 | 1.578 | 1.531 | 1.609 | 1.657 | 1.6405 |
| 4.5 | 1.469 | 1.453 | 1.812 | 1.765 | 1.512 | 1.516 | 1.437 | 1.485 | 1.391 | 1.594 | 1.5494 |
| 5.0 | 1.453 | 1.343 | 1.359 | 1.406 | 1.328 | 1.375 | 1.437 | 1.328 | 1.422 | 1.390 | 1.3841 |
| 5.5 | 1.219 | 1.296 | 1.282 | 1.313 | 1.297 | 1.281 | 1.235 | 1.344 | 1.328 | 1.406 | 1.3001 |

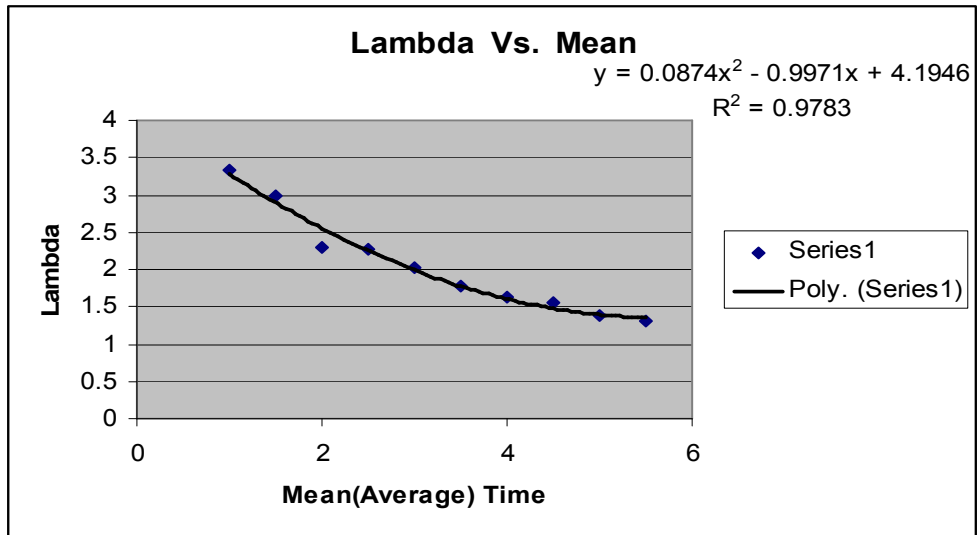

**Figure 2. Mean time in sec versus Lambda ($\lambda$) corresponding to table 2**

**Result:** Above data is best explain by the polynomial of order two and $R^2$=0.9783. Average sorting time clearly depends on the parameter $\lambda$ of Poisson distribution input.

### (c) Geometric Distribution:

Geometric distribution gives the number of failures preceding the first success and has one parameter, namely, the probability of successes P in a trail. We varied *P* from 0.1 to 0.9 for fixed array size n =10,000.

**Table 3. Mean sort time in sec versus probability P of Geometric distribution when n =10,000**

| *P* | TRAIL$_1$ | TRAIL$_2$ | TRAIL$_3$ | TRAIL$_4$ | TRAIL$_5$ | TRAIL$_6$ | MEAN |
|---|---|---|---|---|---|---|---|
| 0.1 | 0.3120 | 0.2500 | 0.2660 | 0.2820 | 0.3130 | 0.2650 | 0.2814 |
| 0.2 | 0.2660 | 0.2340 | 0.2650 | 0.3120 | 0.2500 | 0.3130 | 0.2734 |
| 0.3 | 0.3130 | 0.3120 | 0.2810 | 0.2960 | 0.2660 | 0.2500 | 0.2864 |
| 0.4 | 0.3130 | 0.2970 | 0.2500 | 0.2650 | 0.2810 | 0.2660 | 0.2787 |
| 0.5 | 0.2960 | 0.3120 | 0.2500 | 0.330 | 0.2820 | 0.3120 | 0.2942 |
| 0.6 | 0.3130 | 0.320 | 0.2810 | 0.2500 | 0.2970 | 0.2340 | 0.2812 |
| 0.7 | 0.4060 | 0.4690 | 0.5000 | 0.5160 | 0.3750 | 0.3910 | 0.4428 |
| 0.8 | 0.5160 | 0.5000 | 0.4060 | 0.3590 | 0.3750 | 0.3190 | 0.4245 |
| 0.9 | 0.5320 | 0.5620 | 0.4840 | 0.5940 | 0.5000 | 0.4530 | 0.5208 |

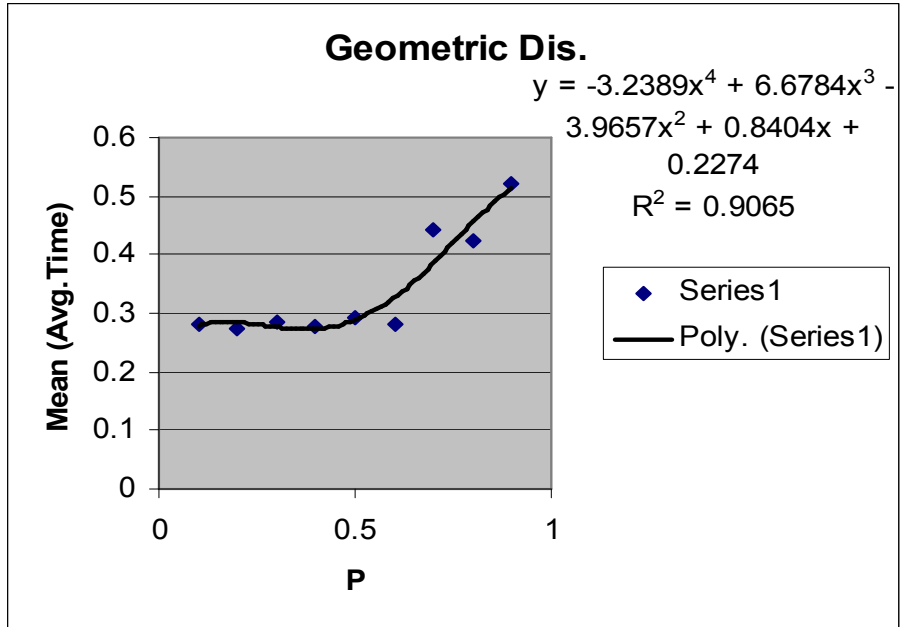

**Figure 3. Mean time in sec versus probability P corresponding to table 3**

**Result:** Above data is best explained by a polynomial of order four and $R^2$=0.9065. Average sorting time clearly depends on the parameter P of Geometric distribution input.

### (d) Continuous Uniform Distribution:

Continuous uniform distribution depends on the parameter of [a, b] where a is minimum value of parameter and b is maximum value parameter. Here take a=0 & b=1 and simulate a U[0, 1] variate and multiplied it with the positive real theta ($\theta$) to generate U[0, $\theta$] variate.

**Table 4. Mean sort time versus theta ($\theta$) where n = 50,000 for U[0, $\theta$] input**

| $\theta$ | Trail$_1$ | Trail$_2$ | Trail$_3$ | Trail$_4$ | Trail$_5$ | Trail$_6$ | Mean |
|---|---|---|---|---|---|---|---|
| **5** | 2.094 | 2.109 | 2.141 | 2.125 | 2.110 | 2.156 | 2.1225 |
| **10** | 1.047 | 1.094 | 1.063 | 1.172 | 1.140 | 1.125 | 1.1069 |
| **15** | 0.719 | 0.75 | 0.734 | 0.766 | 0.782 | 0.796 | 0.758 |
| **20** | 0.562 | 0.578 | 0.531 | 0.641 | 0.547 | 0.625 | 0.581 |
| **25** | 0.469 | 0.5 | 0.453 | 0.531 | 0.468 | 0.516 | 0.489 |
| **30** | 0.437 | 0.391 | 0.422 | 0.407 | 0.438 | 0.406 | 0.417 |
| **35** | 0.36 | 0.344 | 0.359 | 0.375 | 0.391 | 0.343 | 0.362 |
| **40** | 0.328 | 0.313 | 0.344 | 0.421 | 0.39 | 0.32 | 0.353 |
| **45** | 0.297 | 0.329 | 0.281 | 0.312 | 0.344 | 0.343 | 0.315 |
| **50** | 0.266 | 0.281 | 0.297 | 0.312 | 0.25 | 0.329 | 0.313 |

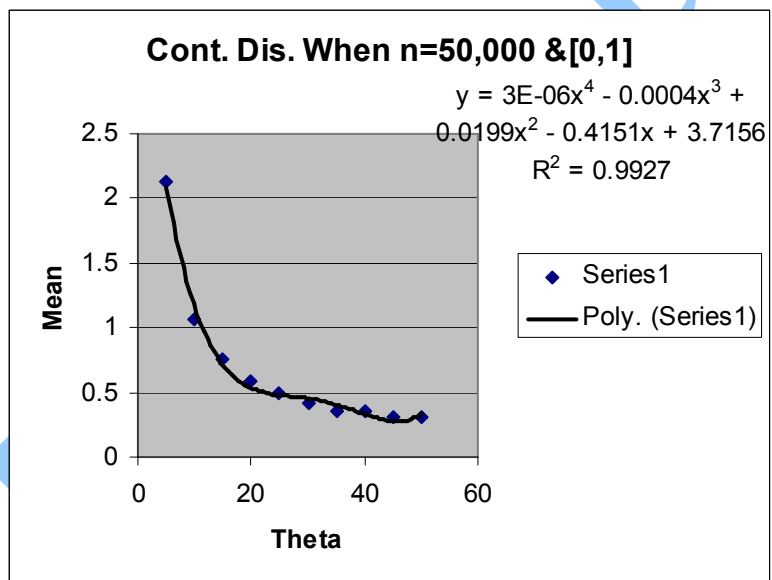

**Figure 4. Mean time versus theta ($\theta$) corresponding to table 4**

**Result:** Above data is explained by a polynomial of order four and $R^2$=0.9927. Average sorting time clearly depends on the parameter $\theta$ (positive real) of Continuous uniform [0, $\theta$] input.

**(e) Exponential distribution:**

This distribution has single parameter, a positive real $\lambda$. For empirical results, we varied the single parameter $\lambda$ and fixed the input array size n = 20,000. The mean of this distribution is $1/\lambda$.

**Table 5. Mean sort time (in sec) versus lambda for Exponential distribution and n= 20,000**

| C(=$\lambda$) | Trail1 | Trail2 | Trail3 | Trail4 | Trail5 | Trail6 | Mean |
|---|---|---|---|---|---|---|---|
| 0.6 | 2.187 | 2.188 | 1.890 | 2.297 | 1.859 | 1.844 | 2.045 |
| 0.7 | 1.844 | 1.828 | 1.781 | 2.016 | 1.750 | 1.875 | 1.849 |
| 0.8 | 1.765 | 1.844 | 1.812 | 2.171 | 1.797 | 1.937 | 1.888 |
| 0.9 | 2.437 | 2.734 | 1.859 | 1.797 | 2.516 | 1.671 | 2.169 |
| 1 | 1.125 | 1.000 | 0.9060 | 0.9220 | 0.9070 | 1.266 | 1.021 |
| 2 | 0.5780 | 0.6720 | 0.6560 | 0.6570 | 0.5630 | 0.5940 | 0.67 |
| 3 | 0.4380 | 0.4540 | 0.4060 | 0.4530 | 0.4370 | 0.4220 | 0.435 |
| 4 | 0.3430 | 0.3280 | 0.3120 | 0.3440 | 0.3290 | 0.3130 | 0.329 |
| 5 | 0.2810 | 0.2660 | 0.2650 | 0.1570 | 0.1560 | 0.2500 | 0.219 |
| 6 | 0.2190 | 0.2100 | 0.2340 | 0.2350 | 0.2180 | 0.1410 | 0.209 |

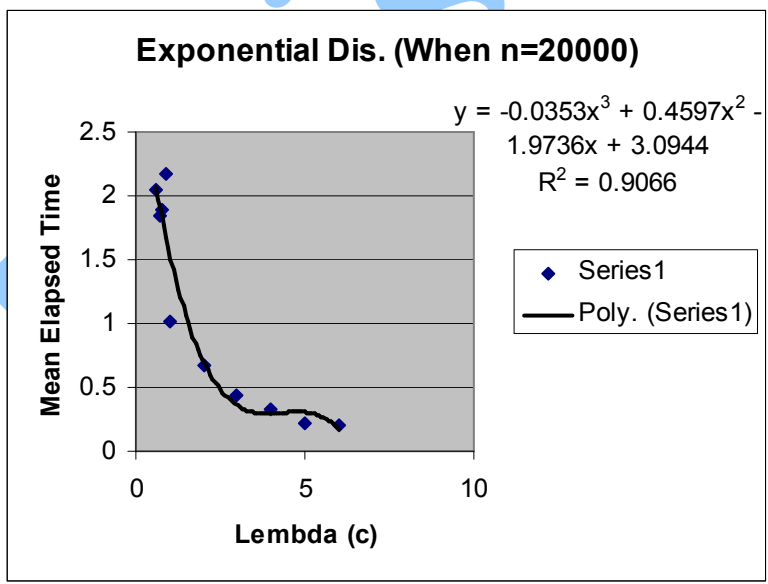

**Figure 5. Mean time (in sec) versus lambda corresponding to table 5**

17

**Result:** Above data is approximately explained by a polynomial of order three and $R^2$=0.9066. Average sorting time clearly depends on the parameter $\lambda$ of Exponential distribution input.

### (f) Normal distribution:

This distribution depends on two parameters mean and variance i.e., ($\mu, \sigma^2$). In this distribution, we have done two experiments; once we kept mean ($\mu$) constant and varied ($\sigma^2$) and vice-versa.

**Case: 1.** We kept a constant variance $\sigma^2$ =100 and varied the mean ($\mu$) with fixed input array size n =20000 and obtained table 6 as follows:

**Table 6. Mean time in sec versus mean ($\mu$)
for Normal (variance = 100 and n=20,000 fixed)**

| $\mu$ | Trail$_1$ | Trail$_2$ | Trail$_3$ | Trail$_4$ | Trail$_5$ | Trail$_6$ | Mean |
|---|---|---|---|---|---|---|---|
| 5 | 0.937 | 0.953 | 1.000 | 0.984 | 1.203 | 1.078 | 1.026 |
| 10 | 0.938 | 1.062 | 1.235 | 1.328 | 1.015 | 0.891 | 1.079 |
| 15 | 1.125 | 0.922 | 0.875 | 1.063 | 1.031 | 1.188 | 1.034 |
| 20 | 1.000 | 1.031 | 0.922 | 1.109 | 1.016 | 1.047 | 1.021 |
| 25 | 0.969 | 1.375 | 1.359 | 1.172 | 0.922 | 1.203 | 1.167 |
| 30 | 1.047 | 1.360 | 1.203 | 1.109 | 0.984 | 0.969 | 1.112 |
| 35 | 0.906 | 1.281 | 1.016 | 1.360 | 1.032 | 1.391 | 1.165 |
| 40 | 1.015 | 1.141 | 1.172 | 1.016 | 1.140 | 1.047 | 1.088 |
| 45 | 0.968 | 0.922 | 1.210 | 1.250 | 1.016 | 1.187 | 1.093 |
| 50 | 1.313 | 1.015 | 0.969 | 1.078 | 0.968 | 0.891 | 1.039 |

**Result:** Above data seems to be random as it is hard to find a definite pattern. Average sorting time probably does not depend on the parameter $\mu$ of Normal distribution input (when variance is fixed).

**Case: 2.** We kept a constant mean $\mu$ =50 and varied the variance ($\sigma^2$) with fixed input array size n =20000 and obtained table 7.

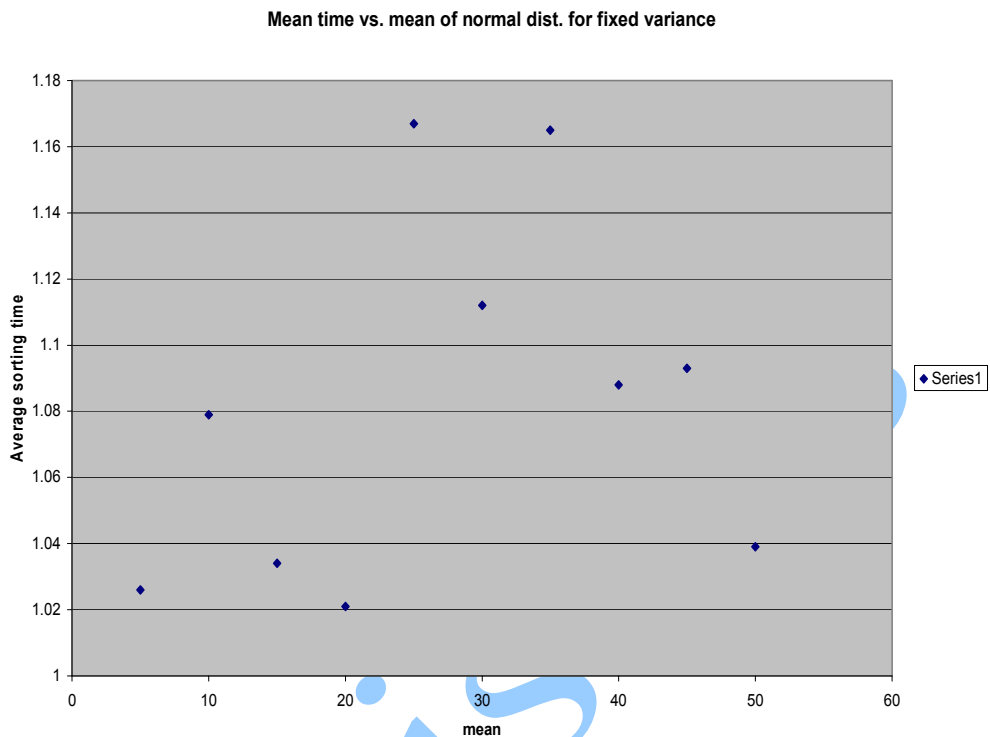

**Figure 6. Mean time in sec versus mean ($\mu$ =m in the program) corresponding to table 6**

**Table 7. Mean time in sec versus variance ($\sigma^2$) for Normal (mean=50 and n=20,000 fixed)**

| $\sigma$ | Trail$_1$ | Trail$_2$ | Trail$_3$ | Trail$_4$ | Trail$_5$ | Trail$_6$ | Mean |
|---|---|---|---|---|---|---|---|
| 10 | 0.984 | 1.203 | 0.969 | 1.031 | 1.016 | 0.954 | 1.193 |
| 20 | 1.172 | 0.953 | 1.125 | 1.219 | 0.985 | 1.171 | 1.105 |
| 30 | 1.407 | 1.218 | 0.937 | 1.015 | 1.031 | 0.891 | 1.084 |
| 40 | 1.047 | 1.203 | 1.031 | 1.094 | 1.375 | 1.062 | 1.136 |
| 50 | 1.109 | 1.188 | 1.137 | 0.968 | 1.235 | 1.203 | 1.180 |
| 60 | 1.031 | 1.313 | 0.968 | 0.984 | 1.375 | 0.953 | 1.104 |
| 70 | 1.313 | 1.375 | 1.128 | 1.390 | 1.297 | 0.891 | 1.248 |
| 80 | 0.937 | 0.984 | 0.969 | 1.093 | 1.187 | 1.000 | 1.029 |
| 90 | 1.375 | 0.894 | 1.000 | 1.297 | 0.953 | 1.187 | 1.133 |
| 100 | 0.938 | 1.297 | 1.141 | 1.031 | 1.016 | 1.375 | 1.133 |

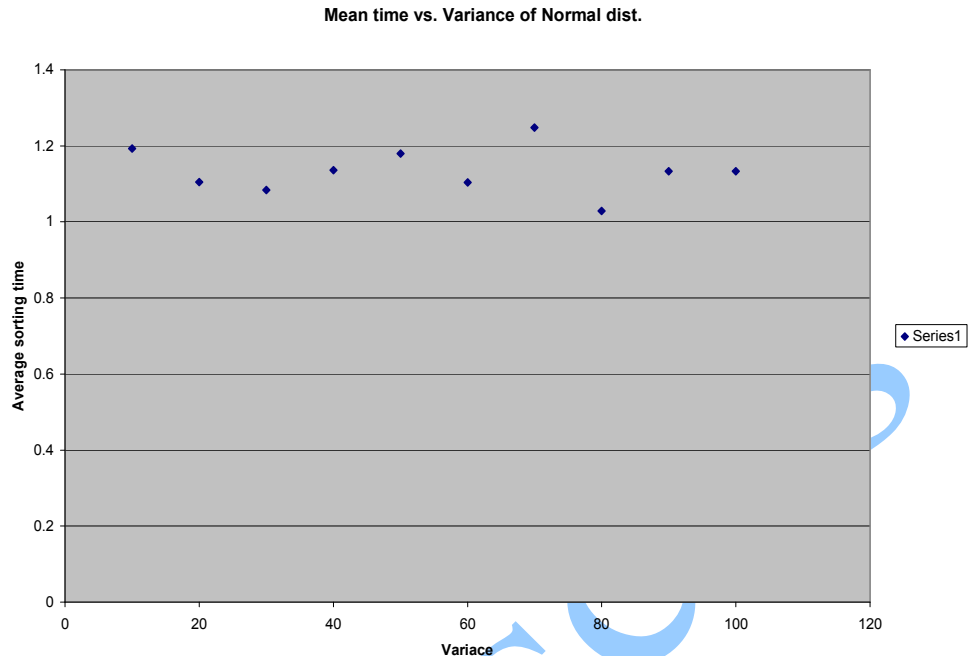

**Figure 7. Mean time in sec versus variance corresponding to table 7**

**Result:** All the values of the average sorting time nearly equal to 1 and does not depend on the variance of Normal distribution input (when mean is fixed).

**System specification:**
    **Processor**    Intel Core Duo @ 667 GHZ
    **Hard Disk**    80 GB
    **RAM**    1 GB
    **O.S.**    Windows XP Professional
    **Software**    **Borland Turbo C ++ Version 5.02**

**Note: -** All empirical results are obtained in Windows XP environment.

## 3. Conclusion and future work

Although it is difficult to establish the exact relationship which the parameters of the input probability distributions characterizing the sorting elements bear on the new sorting algorithm's complexity, it is still possible

20

to statistically investigate some functions that were found to fit the generated data well enough. We are trying to establish some theoretical justifications behind the functional relationships.

Over all empirical results show that in all the distributions (which are discussed) average time complexity depends on the parameters of the input distribution except Normal distribution. Since it was earlier observed by Sourabh, et. al. (2007) that for standard normal input the new sorting algorithm became slowest in the class of probability distributions studied by the authors, we are trying to investigate whether the non-dependence of the time complexity of the sorting algorithm on the parameters for this distribution has any role to play regarding the speed.

## 4. Remarks

1. In a recent work, Khriesat (2007) has found the new sort algorithm to be fast enough to be comparable with some of the faster versions of Quicksort. Khriesat (2007) writes "The new sorting algorithm …is comparable to SedgewickFast, Singleton and Bsort, for values of N ranging from 3000 up to 200,000". Thus although we lose in terms of space complexity, due to the temporary array, there is a definite gain in time by removing the interchanges. This trade off between time and space is nothing uncommon in computer science.

2. Although this paper contains only empirical results it is important for the simple reason that time of an operation is actually its weight and hence we have estimated a weight based bound over a finite range. Such a bound is called a statistical bound and its formal definition alongwith a table of differences between mathematical and statistical complexity bounds can be found in [Chakraborty, Wahi, Sourabh, Saraswati and Mukherjee (2007)]. The ACM review journal Computing Reviews has published at least two reviews revealing this concept [Chakraborty (2008), Chakraborty (2009)].

## 5. Acknowledgement

We thank Professor N. C. Mahanti, Head of the Department of Applied Mathematics, BIT Mesra, Ranchi, India.